\RequirePackage{ifpdf}
\ifpdf 
\documentclass[pdftex]{sigma}
\else
\documentclass{sigma}
\fi

\newcommand{\qquand}{\qquad\text{and}\qquad}
\newcommand{\quand}{\quad\text{and}\quad}
\newcommand{\Real}{\mathbb{R}}
\newcommand{\map}[3]{#1\colon#2\rightarrow#3}
\newcommand{\set}[2]{\left\{\,#1\left.\vphantom{#1#2}\,\right\vert
  \,#2\,\right\}} 
\newcommand{\pai}[2]{\langle\,#1\,,#2\,\rangle}  
\newcommand{\cinfty}[1]{C^\infty(#1)}
\newcommand{\Sec}[2][]{\operatorname{Sec}\nolimits_{#1}(#2)}
\DeclareMathOperator{\id}{id}
\DeclareMathOperator{\pr}{pr}
\newcommand{\pb}{{}^\star}

\newcommand{\pd}[2]{\frac{\partial#1}{\partial#2}}
\newcommand{\at}[1]{\Big\vert_{#1}}

\newcommand{\spC}{^{\scriptscriptstyle\mathsf{C}}}
\newcommand{\spV}{^{\scriptscriptstyle\mathsf{V}}}

\newcommand{\prol}[2][]{\mathcal{T}^{#1}#2}
\newcommand{\TEE}[1][]{\mathcal{T}^E_{#1}E}
\newcommand{\TEP}[1][]{\mathcal{T}^E_{#1}P}
\newcommand{\X}{\mathcal{X}} 
\newcommand{\V}{\mathcal{V}}
\renewcommand{\P}{\mathcal{P}}

\newcommand{\G}{\boldsymbol{G}}
\renewcommand{\t}{\boldsymbol{t}}
\newcommand{\s}{\boldsymbol{s}}
\newcommand{\bepsilon}{\boldsymbol{\epsilon}}

\newcommand{\AJ}[1][E]{\mathcal{A}(J,#1)}
\newcommand{\PJ}[1][E]{\mathcal{P}(J,#1)}
\newcommand{\FL}{\mathcal{F}_L}

\begin{document}

\allowdisplaybreaks

\renewcommand{\PaperNumber}{050}

\renewcommand{\thefootnote}{$\star$}

\FirstPageHeading

\ShortArticleName{Lie Algebroids in Classical Mechanics and Optimal Control}

\ArticleName{Lie Algebroids in Classical Mechanics\\ and Optimal Control\footnote{This paper is a contribution to the Proceedings
of the Workshop on Geometric Aspects of Integ\-rable Systems
 (July 17--19, 2006, University of Coimbra, Portugal).
The full collection is available at
\href{http://www.emis.de/journals/SIGMA/Coimbra2006.html}{http://www.emis.de/journals/SIGMA/Coimbra2006.html}}}

\Author{Eduardo MART{\'I}NEZ}

\AuthorNameForHeading{E. Mart\'{\i}nez}

\Address{Departamento de Matem\'atica Aplicada,
Universidad de Zaragoza,
50009 Zaragoza, Spain}
\Email{\href{mailto:emf@unizar.es}{emf@unizar.es}}
\URLaddress{\url{http://pcmap.unizar.es/~emf/}}

\ArticleDates{Received November 07, 2006, in f\/inal form March
07, 2007; Published online March 20, 2007}

\Abstract{We review some recent results on the theory of Lagrangian systems on Lie algeb\-roids. In particular we consider the symplectic and variational formalism and we study reduction. Finally we also consider optimal control systems on Lie algebroids and we show how to reduce Pontryagin maximum principle.}


\Classification{49S05; 70H25; 22A22; 49J15}

\section{Introduction}

The concept of Lie algebroid was introduced by Pradines in \cite{Pradines1, Pradines2} and has proved to be a useful tool in the formulation and analysis of many problems in dif\/ferential geometry and applied mathematics~\cite{Mackenzie2,CannasWeinstein}.  In the context of geometric Mechanics, a program was proposed by A.\ Weinstein~\cite{Weinstein} in order to develop a theory of  Lagrangian and Hamiltonian systems on Lie algebroids and their discrete analogs on Lie groupoids. In the last years, this program has been actively developed by many authors, and as a result, a powerful mathematical structure is emerging. The purpose of this paper is to review some of such recent developments.

One of the main features of the Lie algebroid framework is its inclusive nature. In what respect to Mechanics, under the same formalism one can describe such disparate situations as Lagrangian systems with symmetry, systems evolving on Lie algebras and semidirect products, or systems with holonomic constraints (see~\cite{LSDLA,SLMCLA} for recent reviews) obtaining in such cases Lagrange--Poincar\'e equations, Poincar\'e equations, Euler--Poincar\'e equations or Euler--Lagrange equations for holonomically constrained problems (see~\cite{CeMaRa,CeMaPeRa}).

While the Lie algebroid approach to Mechanics builds on the geometrical structure of the prolongation of a Lie algebroid~\cite{LMLA}, the origin of Lagrangian Mechanics is the calculus of varia\-tions. It is therefore important to have a variational description of Lagrange's equations for a Lagrangian system def\/ined on a more general Lie algebroid.  We will show that Lagrange's equations for a Lagrangian system on a Lie algebroid are precisely the equations for the critical points of the action functional def\/ined on the set of admissible curves on a Lie algebroid with f\/ixed base endpoints, and we will also show how to f\/ind such equations by means of a Lagrange multiplier method~\cite{VCLA}.

One of the advantages of such a unifying formalism it that morphisms establish relations between these apparently dif\/ferent systems, leading to an adequate way to study reduction theory. In particular we will show how to reduce the variational principle and the symplectic equations in presence of a f\/iberwise surjective morphism of Lie algebroids.

The extension of this ideas to the theory of optimal control systems was initiated in~\cite{ROCT}, and will also be brief\/ly reviewed.  On any Lie algebroid a generalized version of Pontryagin maximum principle can be established in a global and coordinate free way which stresses its geometric properties and can be successfully reduced under morphisms.

There are many other interesting aspects about the application of Lie algebroids to Mechanics which are not covered in this review. For other applications to control theory~\cite{MCSLA}, to discrete mechanics~\cite{DLHMLG} and f\/ield theory~\cite{CFTLAVA,CFTLAMF} see the recent review~\cite{SLMCLA}. For extensions to time-dependent mechanics see~\cite{LASLSAB}.

The paper is organized as follows.  In Section~\ref{preliminaries} we
present some basic facts on Lie algebroids, including results from
dif\/ferential calculus, morphisms and prolongations of Lie algebroids. In
Section~\ref{Mechanics}, we give a brief review of the Hamiltonian and the
Lagrangian formalism of Mechanics on Lie algebroids. In Section~\ref{variational} we show that Lagrange's equation for a Lagrangian system on a~Lie algebroid can be obtained by means of variational calculus by selecting an appropriate class of variations. Much inside is gained by studying the geometry of the inf\/inite dimensional manifold of admissible curves, which is done in Section~\ref{E-paths}. In Section~\ref{reduction} we study the transformation rules induced by  morphism of Lie algebroids on the geometric objects of the theory, and how this is useful in the theory of reduction of Lagrangian systems. Finally in Section~\ref{control} we show how Pontryagin maximum principle can be extended for control systems def\/ined on Lie algebroids and how to reduce optimal control problems.

\section{Preliminaries}
\label{preliminaries}

\subsubsection*{Lie algebroids}
A Lie algebroid structure on a vector bundle $\map{\tau}{E}{M}$ is given by a vector bundle map $\map{\rho}{E}{TM}$ over the identity in $M$, called the anchor, together with a Lie algebra structure on the $\cinfty{M}$-module of sections of $E$ such that the compatibility condition
$
[\sigma,f\eta]=(\rho(\sigma)f)\eta+f[\sigma,\eta]
$
is satisf\/ied for every $f\in\cinfty{M}$ and every $\sigma,\eta\in\Sec{E}$. See~\cite{CannasWeinstein, Mackenzie2} for more information on Lie algebroids.

In what concerns to Mechanics, it is convenient to think of a Lie
algebroid as a generalization of the tangent bundle of $M$. One
regards an element $a$ of $E$ as a generalized velocity, and the
actual velocity $v$ is obtained when applying the anchor to $a$, i.e.,
$v=\rho(a)$. A curve $\map{a}{[t_0,t_1]}{E}$ is said to be
admissible or an $E$-path if $\dot{\gamma}(t)=\rho(a(t))$, where $\gamma(t)=\tau(a(t))$
is the base curve.

A local coordinate system $(x^i)$ in the base manifold $M$ and a local
basis $\{e_\alpha\}$ of sections of~$E$, determine a local coordinate system  $(x^i, y^{\alpha})$ on~$E$.  The anchor and the bracket are locally determined by the local functions $\rho^i_\alpha$ and $C^\alpha_{\beta\gamma}$ on $M$ given by
\begin{gather*}
\rho (e_{\alpha})=\rho _{\alpha}^{i}\frac{\partial}{\partial x^i}
\qquand
[e_{\alpha}, e_{\beta}]=C_{\alpha\beta}^{\gamma}\ e_{\gamma}.
\end{gather*}
The functions $\rho^i_\alpha$ and $C^\alpha_{\beta\gamma}$ satisfy some relations due to the compatibility condition and the Jacobi identity which are called the structure equations:
\begin{gather*}
 \rho_{\alpha}^{j}\frac{\partial \rho _{\beta}^{i}}{\partial
    x^{j}}-\rho _{\beta}^{j}%
  \frac{\partial \rho _{\alpha}^{i}}{\partial x^{j}}=\rho
  _{\gamma}^{i}C_{\alpha\beta}^{\gamma},
  \label{structure.equation.1}
\\
    \rho _{\alpha}^{i}\frac{\partial C_{\beta\gamma}^{\nu}}{\partial x^{i}} +
    \rho _{\beta}^{i}\frac{\partial C_{\gamma\alpha}^{\nu}}{\partial x^{i}} +
    \rho _{\gamma}^{i}\frac{\partial C_{\alpha\beta}^{\nu}}{\partial x^{i}} +
      C_{\beta\gamma}^{\mu}C_{\alpha\mu}^{\nu} +
      C_{\gamma\alpha}^{\mu}C_{\beta\mu}^{\nu} +
      C_{\alpha\beta}^{\mu}C_{\gamma\mu}^{\nu} = 0.
  \label{structure.equation.2}
\end{gather*}

\subsubsection*{Cartan calculus}
The Lie algebroid structure is equivalent to the existence of an exterior dif\/ferential operator  on~$E$, $\map{d}{\Sec{\wedge^kE^*}}{\Sec{\wedge^{k+1}E^*}}$, def\/ined as follows
\begin{gather*}
d \omega(\sigma_0,\dots, \sigma_k)=\sum_{i=0}^{k}
(-1)^i\rho(\sigma_i)(\omega(\sigma_0,\dots,
\widehat{\sigma_i},\dots, \sigma_k))\\
\phantom{d \omega(\sigma_0,\dots, \sigma_k)=}{}+ \sum_{i<j}(-1)^{i+j}\omega([\sigma_i,\sigma_j],\sigma_0,\dots,
\widehat{\sigma_i},\dots,\widehat{\sigma_j},\dots ,\sigma_k),
\end{gather*}
for $\omega\in \Sec{\wedge^k E^*}$ and $\sigma_0,\dots ,\sigma_k\in
\Sec{\tau}$. $d$ is a cohomology operator, that is, $d^2=0$. In
particular, if $\map{f}{M}{\Real}$ is a real smooth function then
$df(\sigma)=\rho(\sigma)f,$ for $\sigma\in \Sec{\tau}$. Locally,
\[
dx^i=\rho^i_{\alpha}e^{\alpha}\qquand de^{\gamma}=-\frac{1}{2}
C^{\gamma}_{\alpha\beta} e^{\alpha}\wedge e^{\beta},
\]
where $\{e^{\alpha}\}$ is the dual basis of $\{e_{\alpha}\}$.  The above mentioned structure equations are but the relations $d^2x^i=0$ and $d^2e^\alpha=0$. We may also def\/ine the Lie derivative with respect to a~section~$\sigma$ of $E$ as the operator $\map{d_\sigma}{\Sec{\wedge^k E^*}}{\Sec{\wedge^k E^*}}$ given by $d_\sigma=i_\sigma\circ d+d\circ i_\sigma$.  Along this paper, except otherwise stated, the symbol $d$ stands for the exterior dif\/ferential on a Lie algebroid.

\subsubsection*{Morphisms}
Given a second Lie algebroid $\map{\tau'}{E'}{M'}$, a vector bundle map $\map{\Phi}{E}{E'}$ over $\map{\varphi}{M}{M'}$ is said to be admissible if it maps admissible curves in $E$ into admissible curves in $E'$, or equivalently if $\rho'\circ\Phi=T\varphi\circ\rho$. The map $\Phi$ is said to be a morphism of Lie algebroids if $\Phi\pb d\theta=d\Phi\pb\theta$ for every $p$-form $\theta\in\Sec{\wedge^pE^*}$. Every morphism is an admissible map.

In coordinates, a vector bundle map $\Phi(x,y)=(\varphi^i(x),\Phi^\alpha_\beta(x)y^\beta)$ is admissible if and only if
\begin{equation*}
\rho'{}^i_\alpha\pd{\varphi^k}{x^i}=\rho^k_\beta\Phi^\beta_\alpha.
\end{equation*}
Moreover, such a map is a morphism if in addition to the above equation it satisf\/ies
\begin{equation*}
\rho^i_\nu\pd{\Phi^\alpha_\mu}{x^i}-\rho^i_\mu\pd{\Phi^\alpha_\nu}{x^i}-
C'{}^\alpha_{\beta\gamma}\Phi^\beta_\mu\Phi^\gamma_\nu=0.
\end{equation*}

\subsubsection*{Prolongation}

In what respect to Mechanics, the tangent bundle to a Lie algebroid, to its dual or to a more general f\/ibration does not have an appropriate Lie algebroid structure. Instead one should use the~so called prolongation bundle which has in every case the appropriate geometrical structures~\cite{LASLSAB,Medina}.

Let $(E, [\ ,\ ], \rho)$ be a Lie
algebroid over a manifold $M$ and $\nu: P \to M$ be a f\/ibration. For every point $p\in P$ we consider the vector space
\[
\TEP[p]=\set{(b,v)\in E_x\times T_pP}{\rho(b)=T_p\nu(v)},
\]
where $T\nu: TP \to TM$ is the tangent map to $\nu$ and
$\nu(p)=x$. The set $\TEP=\cup_{p\in P} \TEP[p]$ has a natural vector bundle structure over $P$, the vector bundle projection $\tau^E_P$ being just the projection $\tau^E_P(b,v)=\tau_P(v)$. We will frequently use the redundant notation $(p,b,v)$ to denote the element $(b,v)\in\TEP[p]$. In this way, the projection $\tau^E_P$ is just the projection onto the f\/irst factor.

The vector bundle $\map{\tau^E_P}{\TEP}{P}$ can be endowed with a Lie algebroid structure. The anchor map is the projection onto the third factor and will also be denoted by $\rho$, that is, the map $\map{\rho}{\TEP}{TP}$ given by $\rho(p,b,v)=v$. To def\/ine the bracket on sections of $\TEP$ we will consider some special sections. A section $Z\in\Sec{\TEP}$ is said to be
projectable if there exists a section $\sigma\in\Sec{E}$ such that $Z(p) =
(p, \sigma(\nu(p)), U(p))$, for all $p \in P$. Now, the bracket of two projectable sections $Z_1$, $Z_2$ given by $Z_i(p)=(p,\sigma_i(\nu(p)), U_i(p))$, $i =1,2$, is given by
\[
[Z_1,Z_2](p)=(p,[\sigma_1,\sigma_2](\nu(p)),[U_1,U_2](p)),
\qquad p \in P.
\]
Since any section of $\TEP$ can be locally written as a $\cinfty{M}$-linear
combination of projectable sections, the def\/inition of the Lie bracket
for arbitrary sections of $\TEP$ follows.

The Lie algebroid $\TEP$ is called the prolongation of $\map{\nu}{P}{M}$ with respect to $E$ or the ${E}$-tangent bundle to $\nu$.

Given local coordinates $(x^i,u^A)$ on $P$ and a local basis
$\{e_\alpha\}$ of sections of $E$, we can def\/ine a local basis
$\{\X_\alpha,\V_A\}$ of sections of $\TEP$ by
\[
\X_\alpha(p)
=\Bigl(p,e_\alpha(\nu(p)),\rho^i_\alpha\pd{}{x^i}\at{p}\Bigr) \qquand
\V_A(p) = \Bigl(p,0,\pd{}{u^A}\at{p}\Bigr).
\]
If $z=(p,b,v)$ is an element of $\TEP$, with $b=z^\alpha
e_\alpha$, then $v$ is of the form $v=\rho^i_\alpha
z^\alpha\pd{}{x^i}+v^A\pd{}{u^A}$, and we can write
\[
z=z^\alpha\X_\alpha(p)+v^A\V_A(p).
\]
Vertical elements are linear combinations of $\{\V_A\}$.

The anchor map $\rho$ applied to a section $Z$ of $\TEP$
with local expression $Z = Z^\alpha\X_\alpha+V^A\V_A$ is the vector
f\/ield on $P$ whose coordinate expression is
\[
\rho(Z) = \rho^i_\alpha Z^\alpha \pd{}{x^i} + V^A\pd{}{u^A}.
\]
The Lie brackets of the elements of the basis are given by
\[
[\X_\alpha,\X_\beta]= C^\gamma_{\alpha\beta}\:\X_\gamma,
\qquad
[\X_\alpha,\V_B]=0
\qquand
[V_A,\V_B]=0,
\]
and, therefore, the exterior dif\/ferential is determined by
\begin{alignat*}{3}
  &dx^i=\rho^i_\alpha \X^\alpha,   &&du^A=\V^A, & \\
  &d\X^\gamma=-\frac{1}{2}C^\gamma_{\alpha\beta}\X^\alpha\wedge\X^\beta,\qquad
  &&d\V^A=0,&
\end{alignat*}
where $\{\X^\alpha,\V^A\}$ is the dual basis to $\{\X_\alpha,\V_A\}$.

\subsubsection*{Prolongation of maps}

We consider now how to prolong maps between two f\/ibrations $\map{\nu}{P}{M}$ and $\map{\nu'}{P'}{M'}$. Let $\map{\Psi}{P}{P'}$ be a map f\/ibered over $\map{\varphi}{M}{M'}$. We consider two Lie algebroids $\map{\tau}{E}{M}$ and $\map{\tau'}{E'}{M'}$ and a map $\map{\Phi}{E}{E'}$ f\/ibered over $\varphi$. If $\Phi$ is admissible, then we can def\/ine a~vector bundle map $\map{\prol[\Phi]{\Psi}}{\TEP}{\prol[E']{P'}}$ by means of
\[
\prol[\Phi]{\Psi}(p,b,v)=(\Psi(p),\Phi(b),T\Psi(v)).
\]
It follows that $\prol[\Phi]{\Psi}$ is also admissible. In~\cite{CFTLAMF} it was proved that $\prol[\Phi]{\Psi}$ is a morphism of Lie algebroids if and only if $\Phi$ is a morphism of Lie algebroids.

In particular, when $E=E'$ and $\Phi=\id$ we have that any map form $P$ to $P'$ f\/ibered over the identity can be prolonged to a morphism $\prol[\id]{\Psi}$ which will be denoted simply by $\prol{\Psi}$.
We will also identify $\prol[E]{M}$ (the prolongation of the `f\/ibration' $\map{\id}{M}{M}$ with respect to $E$) with $E$ itself by means of $(m,b,\rho(b))\equiv b$. With this convention, the projection onto the second factor of $\TEP$ is just $\map{\prol{\nu}}{\TEP}{E}$. It follows that $\prol{\nu}$ is a morphism of Lie algebroids.

\section{Symplectic Mechanics on Lie algebroids}
\label{Mechanics}

By a symplectic structure on a vector bundle
$\map{\pi}{F}{M}$ we mean a section $\omega$ of $\wedge^2F^*$ which is
regular at every point when it is considered as a bilinear form. By a
symplectic structure on a Lie algebroid $E$ we mean a symplectic section $\omega$ of the vector bundle $E$ which is moreover $d$-closed, that is $d\omega=0$. A symplectic Lie algebroid is a  pair $(E,\omega)$ where $E$ is a Lie algebroid and $\omega$ is a symplectic section on it.

On a symplectic Lie algebroid $(E,\omega)$ we can def\/ine a dynamical
system for every function on the base, as in the standard case of a
tangent bundle. Given a function $H\in\cinfty{M}$ there is a~unique
section $\sigma_H\in\Sec{\tau}$ such that
\[
i_{\sigma_H}\omega=dH.
\]
The section $\sigma_H$ is said to be the Hamiltonian section
def\/ined by $H$ and the vector f\/ield $X_H=\rho(\sigma_H)$ is said to be
the Hamiltonian vector f\/ield def\/ined by $H$. In this way we get
the dynamical system $\dot{x}=X_H(x)$.

A symplectic structure $\omega$ on a Lie algebroid $E$ def\/ines a
Poisson bracket $\{\ ,\ \}$ on the base manifold $M$ as
follows. Given two functions $F,G\in\cinfty{M}$ we def\/ine the bracket
\[
\{F,G\}=\omega(\sigma_F,\sigma_G).
\]
It is easy to see that the closure condition $d\omega=0$ implies that
$\{\ ,\ \}$ is a Poisson structure on $M$. In other words, if
we denote by $\Lambda$ the inverse of $\omega$ as bilinear form, then
$\{F,G\}=\Lambda(dF,dG)$. The Hamiltonian dynamical system
associated to $H$ can be written in terms of the Poisson bracket as
$\dot{x}=\{x,H\}$.

Two important particular classes of symplectic dynamical systems on Lie algebroids are the following.

\subsubsection*{Hamiltonian Mechanics~\cite{LSDLA,Medina}} \label{Hamiltonian}
On $\prol[E]{E^*}$, the $E$-tangent to the dual bundle $\map{\pi}{E^*}{M}$, we have a canonical symplectic structure.

The Liouville section $\Theta\in\Sec{(\prol[E]{E^*})^*}$ is the 1-form given by
\begin{equation*}
\label{Lio}
    \pai{\Theta}{(\mu,b,w)}=\pai{\mu}{b}.
\end{equation*}
The canonical symplectic section $\Omega\in \Sec{\wedge^2(\prol[E]{E^*})^*}$ is the dif\/ferential of the Liouville section
\[
\Omega=-d\Theta.
\]
Taking coordinates $(x^i, \mu_{\alpha})$ on $E^*$ and denoting by
$\{\X_{\alpha}, \P^{\beta}\}$ the associated local basis of
sections $\prol[E]{E^*}$, the Liouville and canonical symplectic sections are
written as
\[
\Theta=\mu_{\alpha}\X^{\alpha}\qquand \Omega=\X^{\alpha}\wedge {\mathcal P}_{\alpha} +\frac{1}{2}\mu_{\gamma}
C^{\gamma}_{\alpha\beta}\X^{\alpha}\wedge\X^{\beta} ,
\]
where $\{\X^{\alpha},\P_{\beta}\}$ is the dual
basis of $\{\X_{\alpha},\P^{\beta}\}$.

The Hamiltonian section def\/ined by a function $H\in C^{\infty}(E^*)$ are given in coordinates by
\[
\Gamma_H=\frac{\partial H}{\partial \mu_{\alpha}}\X_{\alpha}-\left( \rho^i_{\alpha} \frac{\partial H}{\partial x^i}+
  \mu_{\gamma} C^{\gamma}_{\alpha\beta} \frac{\partial H}{\partial
    \mu_{\beta}}\right) \P^{\alpha},
\]
and therefore, Hamilton equations are
\begin{equation*}
\label{Hameq}
  \frac{d x^i}{d t}=\rho^i_{\alpha} \frac{\partial H}{\partial
    \mu_{\alpha}},\qquad
   \frac{d\mu_{\alpha}}{d t}=- \rho^i_{\alpha}
  \frac{\partial H}{\partial x^i}- \mu_{\gamma}
  C^{\gamma}_{\alpha\beta} \frac{\partial H}{\partial \mu_{\beta}}.
\end{equation*}

The Poisson bracket $\{\ ,\ \}$ def\/ined by the canonical
symplectic section $\Omega$ on $E^*$ is but the canonical Poisson
bracket, which is known to exists on the dual of a Lie
algebroid~\cite{CannasWeinstein} and Hamilton equations thus coincide with those def\/ined by Weinstein in~\cite{Weinstein}.

\subsubsection*{Lagrangian Mechanics}\label{Lagrangian}
The Lie algebroid approach to Lagrangian Mechanics builds on the geometrical structure of the prolongation of a Lie algebroid~\cite{LMLA} (where one can develop a geometric symplectic treatment of Lagrangian systems parallel to J.\ Klein's formalism~\cite{Klein}).

On the $E$-tangent $\TEE$ to $E$ itself we do not have a canonical symplectic structure. Instead, we have the following two canonical objects: the vertical endomorphism $\map{S}{\TEE}{\TEE}$ which is def\/ined by
  \[
  S(a,b,v)=(a,0,b_a\spV),
  \]
where $b_a\spV$ denotes the vertical lift to $T_aE$ of the element $b\in E$, and
the Liouville section, which is the vertical section corresponding to the Liouville vector f\/ield,
  \[
  \Delta(a)=(a,0,a_a\spV).
  \]

Given a Lagrangian function $L\in\cinfty{E}$ we def\/ine the
Cartan 1-section $\theta_L\in\Sec{(\TEE)^*}$ and the
Cartan 2-section $\omega_L\in\Sec{\wedge^2(\TEE)^*}$ and
the Lagrangian energy $E_L\in C^{\infty}(E)$ as
\begin{equation*}
\label{Cartan-forms}
  \theta_L=S^*(dL) , \qquad
  \omega_L = -d\theta_L\qquand E_L=\mathcal{L}_{\Delta} L-L.
\end{equation*}
If the Cartan 2-section is regular, then it is a symplectic form on the Lie algebroid $\TEE$, and we say that the Lagrangian $L$ is regular. The Hamiltonian section $\Gamma_L$ corresponding to the energy is the Euler--Lagrange section and the equations for the integral curves of the associated vector f\/ield are the Euler--Lagrange equations.

If $(x^i, y^{\alpha})$ are local f\/ibered coordinates on $E$,
$(\rho^i_{\alpha}, C^{\gamma}_{\alpha\beta})$ are the corresponding
local structure functions on $E$ and $\{ \X_{\alpha}, \V_{\alpha}\}$
is the corresponding local basis of sections of $\TEE$ then
$S\X_{\alpha} = \V_{\alpha}$ and $S\V_{\alpha} = 0$, and the Liouville section is $\Delta = y^{\alpha}\V_{\alpha}$. The energy has the expression $E_L=\pd{L}{y^\alpha}y^\alpha-L$,
and the Cartan 2-section is
\begin{equation*}
  \label{omegaL} \omega_L
  =\pd{^2L}{y^\alpha\partial
    y^\beta}\X^\alpha\wedge \V^\beta
  +\frac{1}{2}\left(
    \pd{^2L}{x^i\partial y^\alpha}\rho^i_\beta-\pd{^2L}{x^i\partial
      y^\beta}\rho^i_\alpha+\pd{L}{y^\gamma}C^\gamma_{\alpha\beta}
  \right)\X^\alpha\wedge \X^\beta,
\end{equation*}
from where we deduce that $L$ is regular if and only if the
matrix $\displaystyle{W_{\alpha\beta}=\frac{\partial^2 L}{\partial
    y^{\alpha}\partial y^{\beta}}}$ is regular.
In such case, the local expression of $\Gamma_L$ is
\[
\Gamma_L=y^\alpha\X_\alpha+f^\alpha\V_\alpha ,
\]
where the functions $f^\alpha$ satisfy the linear equations
\begin{equation*}
  \label{free-forces1}
  \pd{^2L}{y^\beta\partial
    y^\alpha}f^\beta+\pd{^2L}{x^i\partial y^\alpha}\rho^i_\beta
  y^\beta +\pd{L}{y^\gamma}C^\gamma_{\alpha\beta}y^\beta
  -\rho^i_\alpha\pd{L}{x^i} =0.
\end{equation*}
Thus, the Euler--Lagrange equations for $L$ are
\begin{equation*}
  \label{Euler-Lagrange}
  \dot{x}^i=\rho_\alpha^iy^\alpha,\qquad
  \frac{d}{dt}\Bigl(\frac{\partial L}{\partial y^\alpha}\Bigr) +
  \frac{\partial L}{\partial y^\gamma}C_{\alpha\beta}^\gamma y^\beta
  -\rho_\alpha^i\frac{\partial L}{\partial x^i} =0.
\end{equation*}

Finally, we mention that, as in the standard case, the relation between the Lagrangian and the Hamiltonian formalism is provided by the Legendre transformation $\map{\FL}{E}{E^*}$ def\/ined by
\begin{equation*}
 \pai{\FL(a)}{b}=\frac{d}{dt}L(a+tb)\big|_{t=0},
\end{equation*}
for $a,b\in E$ with $\tau(a)=\tau(b)$. Then it is easy to see that
\[
\prol{\FL}\pb(\Theta)=\theta_L
\qquand
\prol{\FL}\pb(\Omega)=\omega_L
\]
and therefore, in the regular case, the corresponding Hamiltonian sections are related by $\Gamma_H\circ \FL=\prol{\FL}\circ \Gamma_L$.

\section{Variational description}
\label{variational}

While the Lie algebroid approach to geometric Mechanics builds on the geometrical structure of the $\TEE$, it is well known that the origin of Lagrangian Mechanics is the calculus of variations. Integral curves of a standard Lagrangian system are those tangent lifts of curves on the base manifold which are extremal for the action functional def\/ined on a space of paths.

It is therefore interesting to f\/ind a variational description of Lagrange's equations for a~Lag\-rangian system def\/ined on a more general Lie algebroid. The f\/irst steps in this direction where already done by A.~Weinstein in~\cite{Weinstein} in the case of an integrable Lie algebroid (i.e.\ the Lie algebroid of a Lie groupoid) and by the author in~\cite{Medina,LAGGM}. Finally, a formulation in the inf\/inite dimensional manifold of curves was developed in~\cite{VCLA}.

\subsubsection*{The standard case}
Let us consider f\/irst the situation in the standard case of Lagrangian Mechanics, where $E=TM$. Given a Lagrangian function $L\in\cinfty{TM}$ we want to f\/ind those curves $\map{v}{[t_0,t_1]}{TM}$ which are tangent prolongation of a curve in $M$, that is $v=\dot{\gamma}$ for $\gamma=\tau\circ v$, that connect to given points $m_0$ and $m_1$ in the base manifold $M$, and are extremal points of the action functional $S(v)=\int_{t_0}^{t_1}L(v(t))dt$. One proceed as follows: given a solution $v(t)=\dot{\gamma}(t)$ we consider variations~$v_s(t)$ of~$v(t)$ such that $v_0(t)=v(t)$ and, for every f\/ixed $s$, $v_s(t)=\dot{\gamma}_s(t)$, where $\gamma_s(t)=\tau(v_s(t))$. The inf\/initesimal variation is the vector $Z(t)$ along $v(t)$ given by $Z(t)=\frac{d}{ds} v_s(t)\at{s=0}$, which obviously projects onto the vector f\/ield $W(t)$ along $\gamma(t)$ given by $W(t)=\frac{d}{ds} \gamma_s(t)\at{s=0}$. The Euler--Lagrange equations $\delta L=0$ are derived then by standard manipulations of the condition for stationary points
\[
\frac{d}{ds} S(v_s)\at{s=0}=\int_{t_0}^{t_1}\pai{\delta L}{W}\,dt.
\]

Notice that $\gamma_s$ determines $v_s$, and hence $W$ determines $Z$. This is clear in natural local coordinates $(x^i,v^i)$ on $TM$: we have that $W=W^i(t)\pd{}{x^i}$ and $Z=W^i\pd{}{x^i}+\dot{W}^i\pd{}{v^i}$. In classical notation $\delta x^i= W^i$ and $\delta v^i=\dot{W}^i$, which is but the well known rule for calculating the variation of the velocities as the derivative of the variation of the coordinates
\[
\delta v^i=\delta\left(\frac{dx^i}{dt}\right)=\frac{d}{dt}(\delta x^i)=\dot{W}^i.
\]
Finally notice that due to the f\/ixed endpoints condition we have that $W(t_0)=W(t_1)=0$.

Geometrically, things are a bit more dif\/f\/icult. The vector f\/ield $W(t)$ is a vector f\/ield along the curve $\gamma(t)$, and hence, it is a curve in $TM$ over $\gamma(t)$. If we take the tangent lift, the curve $\dot{W}$ is a curve in $TTM$ over $W(t)$ and therefore def\/ines a vector f\/ield along $W(t)$, instead of a vector f\/ield along $v(t)=\dot{\gamma}(t)$. Therefore, the variation vector f\/ield $Z(t)$ is not just $\dot{W}(t)$ since they are def\/ined at dif\/ferent points. A further operation is needed and this is the so called canonical involution or Tulczyjew involution. It is a map $\map{\chi_{TM}}{TTM}{TTM}$ such that
\[
\chi_{TM}\left(\pd{^2\beta}{s\partial t}(0,0)\right)=\pd{^2\beta}{t\partial s}(0,0),
\]
for every map $\map{\beta}{\Real^2}{M}$ locally def\/ined in a neighborhood of the origin. It follows that the variation vector f\/ield $Z$ is not $\dot{W}$ but it is  \[
Z=\chi_{TM}(\dot{W}).
\]

In many situations this kind of variations is obtained in terms of the f\/lows of vector f\/ields. Given a vector f\/ield $X$ on the manifold $M$ we consider its f\/low $\{\psi_s\}$ and then we def\/ine a~variation of $v(t)$ by $v_s(t)=T\psi_s(v(t))$. It is clear that they are admissible variations, being the base variations $\gamma_s(t)=\psi_s(\gamma(t))$, that $W(t)=X(\gamma(t))$ and $Z(t)=X\spC(v(t))$. In this expression, $X\spC\in\mathfrak{X}(TM)$ is the complete or tangent lift of $X$, whose f\/low is $\{ T\psi_s\}$ and which can be def\/ined in terms of the canonical involution by means of
\[
X\spC(v)=\chi_{TM}(TX(v)),\qquad\text{for all \ \ $v\in TM$.}
\]
Using this kind of variations, the Euler--Lagrange equations can be easily found to be
\[
\frac{d}{dt}\pai{\theta_L}{X\spC}=\mathcal{L}_{X\spC}L,
\]
where $\mathcal L$ denotes the Lie derivative.

\subsubsection*{The general case}
In the general case of a Lagrangian system on an arbitrary Lie algebroid $E$ we can follow a~similar path. We consider a Lagrangian $L\in\cinfty{E}$ and the action
\[
S(a)=\int_{t_0}^{t_1} L(a(t))\,dt
\]
def\/ined on the set of admissible curves on $E$ with f\/ixed base endpoints $m_0$ at $t_0$ and $m_1$ at $t_1$. We look for a variational principle for the Euler--Lagrange equations, that is we have to specify boundary conditions and a class of variations such that the critical points of the action are precisely those curves satisfying Lagrange equations. As we will see such variations are related to complete lifts of sections of $E$.

Every section $\eta$ of $E$ can be naturally lifted to a section of $\TEE$ in two dif\/ferent ways: the vertical lift $\eta\spV$ and the complete lift $\eta\spC$. The structure of Lie algebroid in $\TEE$ is determined by the brackets of such sections,
\[
[\eta\spC,\sigma\spC]=[\sigma,\eta]\spC,
\qquad
[\eta\spC,\sigma\spV]=[\sigma,\eta]\spV
\qquand
[\eta\spV,\sigma\spV]=0.
\]
This relations were used in~\cite{LMLA} to def\/ine the Lie algebroid structure, so that we mimic (and hence extend) the properties of complete and vertical lifts in the tangent bundle, which are on the base for the geometric formalism in the calculus of variations.

In local coordinates, if $\eta=\eta^\alpha e_\alpha$ is a local section of $E$ then the vector f\/ield associated to its complete lift has the local expression
\[
\eta\spC=\eta^\alpha\X_\alpha+
\Bigl(\dot{\eta}^\alpha+C^\alpha_{\beta\gamma}y^\beta\eta^\gamma\Bigr)\V_\alpha,
\]
and the associated vector f\/ield has the expression
\[
\rho(\eta\spC)=\rho^i_\alpha\eta^\alpha\pd{}{x^i}+
\Bigl(\dot{\eta}^\alpha+C^\alpha_{\beta\gamma}y^\beta\eta^\gamma\Bigr)\pd{}{y^\alpha},
\]
where $\dot{f}=\rho^i_\alpha y^\alpha\pd{f}{x^i}$. More generally, one can def\/ine the complete lift of a time-dependent section, which has a similar expression as long as one def\/ines $\dot{f}=\pd{f}{t}+\rho^i_\alpha y^\alpha\pd{f}{x^i}$.

Using the properties of complete and vertical lifts, it is easy to see that the Euler--Lagrange equations $i_{\Gamma_L}\omega_L=d E_L$ can also be written in the form
\[
d_{\Gamma_L}\pai{\theta_L}{\sigma\spC}=d_{\sigma\spC}L,
\]
for every time-dependent section $\sigma$ of $E$. From this expression one can deduce that the inf\/i\-ni\-te\-si\-mal variations that one must consider are precisely the vector f\/ield associated to the complete lifts of sections of $E$.

The above observation is not only a formal statement, but can be carried on precisely in terms of the f\/low associated to a time-dependent section (see~\cite{LAGGM, Medina}). For simplicity in the exposition I will consider only time-independent sections. If $\eta$ is a section of $E$ then the f\/low~$\Phi_s$ of the vector f\/ield $\rho(\eta\spC)\in\mathfrak{X}(E)$ projects to the f\/low $\varphi_s$ of the vector f\/ield $\rho(\eta)\in\mathfrak{X}(M)$. For every f\/ixed $s$, the map $\Phi_s$ is a vector bundle map which is a morphism of Lie algebroids over~$\varphi_s$. The pair $(\Phi_s,\varphi_s)$ is said to be the f\/low of the section $\eta\in\Sec{E}$, and we have that
\[
d_\eta\theta=\frac{d}{ds}\Phi\pb_s\theta\at{s=0},
\]
for every tensor f\/ield $\theta$ over $E$.

Given an admissible curve $a(t)$ we consider a section $\sigma$ of $E$ and its f\/low $(\Phi_s,\varphi_s)$ the variations $a_s(t)=\Phi_s(a(t))$, which are also admissible curves, since $\Phi_s$ are morphisms of Lie algebroids. If we moreover consider sections $\eta$ vanishing at the endpoints, $\eta(m_0)=\eta(m_1)=0$, then the varied curve has f\/ixed endpoints, $\tau(a_s(t_0))=m_0$ and $\tau(a_s(t_1))=m_0$.

Notice that, in general, there are more general variations preserving the admissibility of curves than those considered here. Nevertheless, we have to chose exactly the ones we have chosen: if we consider a restricted class of  variations, we will get unspecif\/ied dynamics and if we consider a more general class of variations we will get some constraints. One can clearly see this fact in the case of a Lie algebra, where every curve is admissible and hence every variation preserves admissible curves.

\subsubsection*{The canonical involution}
In the argument given above, in order to def\/ine a variation we need a section of $E$ def\/ined in a~neighborhood of the base path. As in the case of the standard Lagrangian mechanics, a~dif\/ferent procedure consists in using the canonical involution for def\/ining variations.

 Indeed, the canonical involution can also be def\/ined on any Lie algebroid $E$ (see~\cite{LSDLA} for the details). That is, there exists a canonical map $\map{\chi_E}{\TEE}{\TEE}$ such that $\chi_E^2=\id$ and it is  def\/ined by $\chi_E(a,b,v)=(b,a,\bar{v})$, for every $(a,b,v)\in\TEE$, where $\bar{v}\in T_bE$ is the vector which projects to $\rho(a)$ and satisf\/ies
\[
\bar{v}\hat{\theta}=v\hat{\theta}+d\theta(a,b)
\]
for every section $\theta$ of $E^*$, where $\hat{\theta}\in\cinfty{E}$ is the linear function associated to $\theta$. In local coordinates
the canonical involution is given by
\[
\chi_E(x^i,y^\alpha,z^\alpha,v^\alpha)=(x^i,z^\alpha,y^\alpha,v^\alpha+C^\alpha_{\beta\gamma}z^\beta y^\gamma).
\]
From this expression is clear that the complete lift of a section $\eta\in\Sec{E}$ can be given in terms of the canonical involution by
\[
\eta\spC(a)=\chi_E\bigl(\prol{\eta}(a)\bigr)
\qquad\text{for all \ \ $a\in E$.}
\]

This formula suggests to consider the following map. Given an admissible curve $\map{a}{\Real}{E}$ over $\gamma=\tau\circ a$ we consider the map $\Xi_a$ from sections of $E$ along $\gamma$ to sections of $TE$ along $a$, i.e.\ $\map{\Xi_a}{\Sec[\gamma]{E}}{\Sec[a]{TE}}$, given by
\[
\Xi_a(\sigma)=\rho^1(\chi_E(\sigma,a,\dot{\sigma})).
\]
The local expression of the map $\Xi_a$ is
\[
\Xi_a(\sigma)(t)=\rho^i_\alpha(\gamma(t))\sigma^\alpha(t)\pd{}{x^i}\at{a(t)}
+\Bigl(\dot{\sigma}^\alpha(t)+C^\alpha_{\beta\gamma}(\gamma(t))a^\beta(t)\sigma^\gamma(t)\Bigr)\pd{}{y^\alpha}\at{a(t)},
\]
where $a$ and $\sigma$ have the local expression $a(t)=(\gamma^i(t),a^\alpha(t))$ and $\sigma(t)=(\gamma^i(t),\sigma^\alpha(t))$.

\section[The manifold of $E$-paths]{The manifold of $\boldsymbol{E}$-paths}
\label{E-paths}

To get some more insight into the variational principle that we have obtained, we can analyze the situation from the point of view of the geometry of the inf\/inite dimensional manifold of admissible curves.

\subsubsection*{Homotopy of $\boldsymbol{E}$-paths}

Let $I=[0,1]$ and $J=[t_0,t_1]$, and denote the coordinates in $\Real^2$ by $(s,t)$.
Given a vector bundle map $\map{\Phi}{T\Real^2}{E}$, denote $a(s,t)=\Phi(\partial_t|_{(s,t)})$ and $b(s,t)=\Phi(\partial_s|_{(s,t)})$, so that we can write $\Phi=adt+bds$.

\begin{definition}
Two $E$-paths $a_0$ and $a_1$ are said to be $E$-homotopic if there exists a morphism  of Lie algebroids $\map{\Phi}{TI\times TJ}{E}$, $\Phi=a dt+bds$, such that
\begin{alignat*}{3}
&a(0,t)=a_0(t),\qquad
&&b(s,t_0)=0,&\\
&a(1,t)=a_1(t),\qquad
&&b(s,t_1)=0.&
\end{alignat*}
We will say that $\Phi$ is an $E$-homotopy from the $E$-path $a_0$ to the $E$-path $a_1$.
\end{definition}

It follows that the base map is a homotopy (in the usual sense) with f\/ixed endpoints between the base paths. Notice that $a(s,t)$ is a variation of $a(0,t)$ and one should think of $b(s,t)$ as the vector generating the variation.

\begin{theorem}[\cite{Rui}]
The set of $E$-paths
\[
\AJ=\set{\map{a}{J}{E}}{\rho\circ a=\frac{d}{dt}(\tau\circ a)}
\]
is a Banach submanifold of the Banach manifold of $C^1$-paths whose base path is $C^2$.
Every $E$-homotopy class is a smooth Banach manifold and the partition into equivalence classes is a~smooth foliation. The distribution tangent to that foliation is given by $a\in\AJ \mapsto F_a$ where
\[
F_a=\set{\Xi_a(\sigma)\in T_a\AJ}{\sigma(t_0)=0\quand \sigma(t_1)=0}.
\]
and the codimension of $F$ is equal to $\dim(E)$. The $E$-homotopy equivalence relation is regular if and only if the Lie algebroid is integrable (i.e.\ it is the Lie algebroid of a Lie groupoid).
\end{theorem}

\subsubsection*{The space of $\boldsymbol{E}$-paths}

Therefore, on the same set $\AJ$ there are two natural dif\/ferential manifold structures: as a~submanifold of the set of $C^1$ paths in $E$, which will be denoted just $\AJ$, and the structure induced by the foliation into $E$-homotopy classes, which will be denoted $\PJ$. We will refer to it as the space of $E$-paths on the Lie algebroid $E$. The structure of $\AJ$ is relevant when one wants to study the relation between neighbor $E$-homotopy classes, as it is the case in the problem of integrability of Lie algebroids to Lie groupoids. The structure of $\PJ$ is just the structure that one needs in Mechanics, where one does not have the possibility to jump from one $E$-homotopy class to another.

Notice that every homotopy class is a connected component of $\PJ$, and the identity def\/ines a smooth map $\map{i}{\PJ}{\AJ}$ which is an (invertible) injective immersion. The image by $i$ of a leaf is an immersed (in general not embedded) submanifold of $\AJ$. The tangent space to $\PJ$ at $a$ is $T_a\PJ=F_a$. The topology of $\PJ$ is f\/iner than the topology of $\AJ$. In particular, if $\map{G}{\AJ}{Y}$ is a smooth map, then $\map{G\circ i}{\PJ}{Y}$ is also smooth.

\subsubsection*{Variational description}

With the manifold structure that we have previously def\/ined on the space of $E$-paths, we can formulate the variational principle in a standard way. Let us f\/ix two points $m_0,m_1\in M$ and consider the set $\PJ_{m_0}^{m_1}$ of those $E$-paths with f\/ixed base endpoints equal to $m_0$ and $m_1$, that~is
\[
\PJ_{m_0}^{m_1}=\set{a\in\PJ}{\tau(a(t_0))=m_0\quand \tau(a(t_1))=m_1}.
\]
We remark that $\PJ_{m_0}^{m_1}$ is a Banach submanifold of $\PJ$, since it is a disjoint union of Banach submanifolds (the $E$-homotopy classes of curves with base path connecting such points). On the contrary, there is no guaranty that the analog set $\AJ_{m_0}^{m_1}$ is a manifold (see~\cite{PiTa}).

\begin{theorem}[\cite{VCLA}]
Let $L\in\cinfty{E}$ be a Lagrangian function on the Lie algebroid $E$ and fix two points $m_0, m_1\in M$. Consider the action functional $\map{S}{\PJ}{\Real}$ given by $S(a)=\int_{t_0}^{t_1} L(a(t))dt$. The critical points of $S$ on the Banach manifold $\PJ_{m_0}^{m_1}$ are precisely those elements of that space which satisfy Lagrange's equations.
\end{theorem}

\subsubsection*{Lagrange Multipliers}
\label{LagrangeMultipliers}
We can also analyze the problem by using Lagrange multipliers method by imposing a condition on $\AJ$ which represents the constraint that our $E$-paths are in a given $E$-homotopy class. This is connected with the theory of \textsl{Lin constraints}~\cite{CeIbMa}.

We consider only the case of an integrable Lie algebroid, since in the contrary we will not have a dif\/ferential manifold structure in the set of $E$-homotopy equivalence classes. In this case, the foliation def\/ined by the $E$-homotopy equivalence relation is a regular foliation so that quotient $\G=\AJ/\sim$ has the structure of quotient manifold and the quotient projection $\map{q}{\AJ}{\G}$ is a submersion. Def\/ining the source and target maps by $\s([a])=\tau(a(t_0))$ and $\t([a])=\tau(a(t_1))$, the unit map $\map{\bepsilon}{M}{\G}$ by $\bepsilon(m)=[0_m]$, where $0_m$ denotes the constant curve with value $0\in E_m$, and the multiplication induced by concatenation of $E$-paths, we have that $\G$ is the source simply-connected Lie groupoid with Lie algebroid $E$. See~\cite{Rui} for the details.

Given $g\in\G$, we can select the curves in an $E$-homotopy class as the set $q^{-1}(g)$. Therefore we look for the critical points of the functional $S(a)=\int_{t_0}^{t_1} L(a(t))\,dt$ def\/ined in $\AJ$, constrained by the condition $q(a)=g$. Since $q$ is a submersion, there are not singular curves for the constraint map, and we can use Lagrange multiplier method in the standard form~\cite{MTA}.

\begin{theorem}[\cite{VCLA}]
Let $\map{S}{\AJ}{\Real}$, be the action functional $S(a)=\int_{t_0}^{t_1} L(a(t))\,dt$. An admissible curve $a\in\AJ$ is a solution of Lagrange's equations if and only if there exists $\mu\in T_g^*\G$ such that $dS(a)=\mu\circ T_aq$.
\end{theorem}

A more convenient setting for this constrained problem consists in f\/ixing one of the endpoints. Given $m_0\in M$, the subset $\AJ_{m_0}$ of those $E$-paths whose base path start at $m_0$, $\AJ_{m_0}=\set{a\in\AJ}{\tau(a(t_0))=m_0}$,
is a smooth Banach submanifold of $\AJ$. On $\AJ_{m_0}$ we def\/ine the map $\map{p}{\AJ_{m_0}}{\s^{-1}(m_1)}$ by $p(a)=L_{g^{-1}}(q(a))$. With the help of this map, the constraint reads $p(a)=\bepsilon(m_1)$, because an $E$-path is in $q^{-1}(g)$ if and only if it is in  $p^{-1}(\bepsilon(m_1))$. Then the tangent space to $\AJ_{m_0}$ at $a\in\AJ_{m_0}$ is
\[
T_a\AJ_{m_0}=\set{\Xi_a(\sigma)}{\sigma(t_0)=0}.
\]
$p$ is a submersion and the tangent map $\map{T_ap}{T_a\AJ_{m_0}}{E_{m_1}}$ to $p$ at a point $a\in p^{-1}(\bepsilon(m_1))$, is given by the endpoint mapping
\[
T_ap(\Xi_a(\sigma))=\sigma(t_1)
\]
for every $\sigma\in\Sec[\gamma]{E}$ such that $\sigma(t_0)=0$. If we now apply Lagrange multiplier theorem we obtain the following result.

\begin{theorem}[\cite{VCLA}]
Let $S_{m_0}$ be the restriction of the action functional to the submanifold $\AJ_{m_0}$. An admissible curve $a\in\AJ_{m_0}$ is a solution of Lagrange's equations if and only if it there exists $\lambda\in E^*_{m_1}$ such that  $dS_{m_0}(a)=\lambda\circ T_aq$. The multiplier $\lambda$ is given explicitly by $\lambda=\theta_L(a(t_1))$.
\end{theorem}

\section{Morphisms and reduction}
\label{reduction}

One important advantage of dealing with Lagrangian systems evolving on
Lie algebroids is that the reduction procedure can be naturally
handled by considering morphisms of Lie algebroids, as it was already
observed by Weinstein~\cite{Weinstein}. We study in this section the
transformation laws of the dif\/ferent geometric objects in our theory
and we apply these results to the study of the reduction theory.

\subsubsection*{Mappings induced by morphisms}

We recall that admissible maps are precisely those maps which transforms admissible curves into admissible curves. Therefore an admissible map $\map{\Phi}{E}{E'}$ induces a map between $E$-paths by composition $a\mapsto \Phi\circ a$. We prove now that such a map is smooth provided that $\Phi$ is a morphism.

More precisely, let $\map{\Phi}{E}{E'}$ be an admissible map. It is easy to see that, $\Phi$ is a Lie algebroid morphism if and only if $T{\Phi}\circ\Xi_a(\sigma)=\Xi_{\Phi\circ a}(\Phi\circ\sigma)$ for every $E$-path $a$ and every section $\sigma$ along the base path $\tau\circ a$. It follows that morphisms transform vectors tangent to the foliation into vectors tangent to the foliation, and hence they induce a smooth map between path spaces.

\begin{proposition}[\cite{VCLA}]
Given a morphism of Lie algebroids $\map{\Phi}{E}{E'}$ the induced map  $\map{\hat{\Phi}}{\PJ}{\PJ[E']}$ given by $\hat{\Phi}(a)=\Phi\circ a$ is smooth.
Moreover,
\begin{itemize}\itemsep=0pt
\item If $\Phi$ is fiberwise surjective then $\hat{\Phi}$ is a submersion.
\item If $\Phi$ is fiberwise injective then $\hat{\Phi}$ is a immersion.
\end{itemize}
\end{proposition}

As a consequence, the variational structure of the problem is not broken by reduction. On the contrary, reduction being a morphism of Lie algebroids, preserves such structure. The above results says that morphisms transforms admissible variations into admissible variations. Therefore, a morphism induces relations between critical points of functions def\/ined on path spaces, in particular between the solution of Lagrange's equations.

\subsubsection*{Reduction of the variational principle}
Consider a morphism $\map{\Phi}{E}{E'}$ of Lie algebroids and the induced map between the spaces of paths $\map{\hat{\Phi}}{\PJ}{\PJ[E']}$. Consider a Lagrangian $L$ on $E$ and a Lagrangian $L'$ on $E'$ which are related by $\Phi$, that is, $L=L'\circ\Phi$. Then the associated action functionals $S$ on~$\PJ$ and~$S'$ on~$\PJ[E']$ are related by $\hat{\Phi}$, that is  $S'\circ \hat{\Phi}=S$. Indeed,
\[
S'(\hat{\Phi}(a))=S'(\Phi\circ a)
=\int_{t_0}^{t_1}(L'\circ\Phi\circ a)(t)\,dt
=\int_{t_0}^{t_1}(L\circ a)(t)\,dt
=S(a).
\]

The following result is already in~\cite{Weinstein} but the proof is dif\/ferent.
\begin{theorem}[\cite{Weinstein, VCLA}]
\label{reconstruction}
Let $\map{\Phi}{E}{E'}$ be a morphism of Lie algebroids. Consider a Lagran\-gian~$L$ on $E$ and a Lagrangian $L'$ on $E'$ such that $L=L'\circ\Phi$. If $a$ is an $E$-path and $a'=\Phi\circ a$ is a~solution of Lagrange's equations for $L'$ then $a$ itself is a solution of Lagrange's equations for~$L$.
\end{theorem}
\begin{proof}
Since $S'\circ \hat{\Phi}=S$ we have that $\pai{dS'(\hat{\Phi}(a))}{T_a\hat{\Phi}(v)} =\pai{dS(a)}{v}$ for every $v\in T_a\PJ_{m_0}^{m_1}$. If $\hat{\Phi}(a)$ is a solution of Lagrange's equations for $L'$ then $dS'(\hat{\Phi}(a))=0$, from where it follows that $dS(a)=0$.
\end{proof}

From the above relations between the action functionals it readily follows a reduction theorem.

\begin{theorem}[\cite{VCLA}]
Let $\map{\Phi}{E}{E'}$ be a fiberwise surjective morphism of Lie algebroids. Consider a Lagrangian $L$ on $E$ and a Lagrangian $L'$ on $E'$ such that $L=L'\circ\Phi$. If $a$ is a solution of Lagrange's equations for $L$ then $a'=\Phi\circ a$ is a solution of Lagrange's equations for $L'$.
\end{theorem}
\begin{proof}
Since $S'\circ \hat{\Phi}=S$ we have that $\pai{dS'(\hat{\Phi}(a))}{T_a\hat{\Phi}(v)} =\pai{dS(a)}{v}$ for every $v\in T_a\PJ_{m_0}^{m_1}$. If $\Phi$ is f\/iberwise surjective, then $\hat{\Phi}$ is a submersion, from where it follows that $\hat{\Phi}$ maps critical points of $S$ into critical points of $S'$, i.e.\ solutions of Lagrange's  equations for $L$ into solutions of Lagrange's equations for $L'$.
\end{proof}

\subsubsection*{Reduction of the symplectic form and the dynamics}

Reduction can also be studied in the context of the symplectic formalism on Lie algebroids, see~\cite{NHLSLA} or~\cite{SLMCLA} for the details.

\begin{proposition}\label{transformation-S}
  Let $\map{\Phi}{E}{E'}$ be a morphism of Lie algebroids, and
  consider the $\Phi$-tangent prolongation of $\Phi$, i.e
  $\map{\prol[\Phi]{\Phi}}{\TEE}{\prol[E']{E'}}$.  Let $S$ and $S'$, and $\Delta$ and $\Delta'$, be the the vertical endomorphisms and the Liouville
  sections on $E$ and $E'$, respectively. Then,
  \[
  \prol[\Phi]{\Phi}\circ\Delta = \Delta'\circ\Phi
  \qquand
  \prol[\Phi]{\Phi}\circ S = S'\circ \prol[\Phi]{\Phi}.
  \]
\end{proposition}

\begin{proposition}\label{transformation-omegaL}
  Let $L\in\cinfty{E}$ be a Lagrangian function, $\theta_L$ the Cartan
  form and $\omega_L=-d\theta_L$.  Let $\map{\Phi}{E}{E'}$ be a Lie
  algebroid morphism and suppose that $L=L'\circ \Phi$, with $L'\in
  \cinfty{E'}$ a~Lagrangian function.  Then, we have
  \[
  (\prol[\Phi]{\Phi})\pb\theta_{L'}=\theta_L,
  \qquad
  (\prol[\Phi]{\Phi})\pb\omega_{L'}=\omega_L
  \qquand
  (\prol[\Phi]{\Phi})\pb E_{L'}=E_L.
  \]
\end{proposition}

The transformation of the symplectic equation is easily found by means of standard arguments, and we f\/ind
\[
(\prol[\Phi]{\Phi})\pb\bigl(i_{\Gamma_{L'}}\omega_{L'}-dE_{L'}\bigr)-
(i_{\Gamma_L}\omega_L-dE_L)
    =\omega_{L'}\bigl(\Gamma_{L'}\circ\Phi-\prol[\Phi]{\Phi} \circ
    \Gamma_L,\prol[\Phi]{\Phi}(\,\,\cdot\,\,)\bigr),
\]
It follows that, if $\Phi$ is a f\/iberwise surjective morphism and $L$ is a regular Lagrangian on $E$, then $L'$ is a regular Lagrangian on $E'$ (note that $\map{\prol[\Phi]{\Phi}}{\TEE}{\prol[E']{E'}}$ is a
f\/iberwise surjective morphism) we have that the dynamics of both systems is
uniquely def\/ined, and it is related as follows.

\begin{theorem}
  Let $\Gamma_L$ and
  $\Gamma_{L'}$ be the solutions of the dynamics defined by the
  Lagrangians $L$ and~$L'$, respectively, with $L=L'\circ\Phi$. If $\Phi$
  is a fiberwise surjective morphism and $L$ is a regular Lagrangian, then
  $L'$ is also a regular Lagrangian and
  \[
  \prol[\Phi]{\Phi}\circ\Gamma_L=\Gamma_{L'}\circ\Phi.
  \]
\end{theorem}

Finally, by introducing some constraints one can also study nonholonomic mechanical systems on Lie algebroids. See~\cite{NHLSLA, SLMCLA} for the general theory and results on reduction for nonholonomic systems.

\subsubsection*{Examples} We present here some examples where the reduction process indicated above can be applied. See~\cite{VCLA,NHLSLA,SLMCLA} for more examples.

{\bf Lie groups.}
Consider a Lie group $G$ and its Lie algebra $\mathfrak{g}$. The map $\map{\Phi}{TG}{\mathfrak{g}}$ given by $\Phi(g,\dot{g})=g^{-1}\dot{g}$ is a morphism of Lie algebroids, which is f\/iberwise bijective. As a consequence if $L$ is a left-invariant Lagrangian function on $TG$ and $L'$ is the projected Lagrangian on the Lie algebra $\mathfrak{g}$, that is $L(g,\dot{g})=L'(g^{-1}\dot{g})$, then every solution of Lagrange's equations for $L$ projects by $\Phi$ to a solution of  Lagrange's equations for $L'$. Moreover, since $\Phi$ is surjective every solution can be found in this way: if the projection $\xi(t)=g(t)^{-1}\dot{g}(t)$ of an admissible curve $(g(t),\dot{g}(t))$ is a solution of $L'$, then $(g(t),\dot{g}(t))$ is a solution for $L$. Thus, the Euler--Lagrange equations on the group reduce to the Euler--Poincar\'e equations on the Lie algebra, both being symplectic equations in the Lie algebroid sense.

{\bf Lie groupoids.}
Consider a Lie groupoid $\G$ over $M$ with source $\s$ and target $\t$,  and with Lie algebroid $E$. Denote by $T^{\s}\G\to\G$ the kernel of $T\s$ with the structure of Lie algebroid as integrable subbundle of $T\G$. Then the map $\map{\Phi}{T^{\s}\G}{E}$ given by left translation to the identity, $\Phi(v_g)=TL_{g^{-1}}(v_g)$ is a morphism of Lie algebroids, which is moreover f\/iberwise surjective. As a consequence, if $L$ is a Lagrangian function on $E$ and  $\boldsymbol{L}$ is the associated left invariant Lagrangian on $T^{\s}\G$, then the solutions of Lagrange's equations for $\boldsymbol{L}$ project by $\Phi$ to solutions of the Lagrange's equations. Since $\Phi$ is moreover surjective, every solution can be found in this way.

{\bf Group actions.}
We consider a Lie group $G$ acting free and properly on a manifold $Q$, so that the quotient map $\map{\pi}{Q}{M}$ is a principal bundle. We consider the standard Lie algebroid structure on $E=TQ$ and the associated Atiyah algebroid $E'=TQ/G\to M$. The quotient map $\map{\Phi}{E}{E'}$, $\Phi(v)=[v]$ is a Lie algebroid morphism and it is f\/iberwise bijective. Every $G$-invariant Lagrangian on $TQ$ def\/ines uniquely a Lagrangian $L'$ on $E'$ such that $L'\circ\Phi=L$. Therefore every solution of the $G$-invariant Lagrangian on $TQ$ projects to a solution of the reduced Lagrangian on $TQ/G$, and every solution on the reduced space can be obtained in this way. Thus, the Euler--Lagrange equations on the principal bundle reduce to the Lagrange--Poincar\'e equations on the Atiyah algebroid,  both being symplectic equations in the Lie algebroid sense.

\section{Optimal control theory}
\label{control}

As it is well known, optimal control theory is a generalization of classical mechanics. It is therefore natural to see whether our results can be extended to this more general context. The central result in the theory of optimal control systems is Pontryagin maximum principle. The reduction of optimal control problems can be performed within the framework of Lie algebroids, see~\cite{ROCT}. This was done as in the case of classical mechanics, by introducing a general principle for any Lie algebroid and later studying the behavior under morphisms of Lie algebroids.

\subsubsection*{Pontryagin maximum principle~\cite{ROCT}}

By a control system on a Lie algebroid $\map{\tau}{E}{M}$ with control space $\map{\pi}{B}{M}$ we mean a~section $\sigma$ of $E$ along $\pi$. A trajectory of the system $\sigma$ is an integral curve of the vector f\/ield~$\rho(\sigma)$ along $\pi$. Given an index function $L\in\cinfty{B}$ we want to minimize the integral of~$L$ over some set of trajectories of the system which satisf\/ies some boundary conditions. Then we def\/ine the Hamiltonian function $H\in\cinfty{E^*\times_MB}$ by $H(\mu,u)=\pai{\mu}{\sigma(u)}-L(u)$ and the associated Hamiltonian control system $\sigma_H$ (a section of $\prol[E]{E^*}$ along $\map{\mathrm{\pr}_1}{E^*\times_MB}{E^*}$) def\/ined on a subset of the manifold $E^*\times_M B$, by means of the symplectic equation
\begin{equation}\tag{$\star$}
\label{pontryagin}
i_{\sigma_H}\Omega=dH.
\end{equation}
The integral curves of the vector f\/ield $\rho(\sigma_H)$ are said to be the critical trajectories.

In the above expression, the meaning of $i_{\sigma_H}$ is as follows. Let $\map{\Phi}{E}{E'}$ be a  morphism over a map $\map{\varphi}{M}{M'}$ and let $\eta$ be a section of $E'$ along $\varphi$. If $\omega$ is a section of $\bigwedge^pE'{}^*$ then $i_\eta\omega$ is the section of $\bigwedge^{p-1}E^*$ given by
\[
(i_\eta\omega)_m(a_1,\ldots,a_{p-1})=\omega_{\varphi(n)}(\eta(m),\Phi(a_1),\ldots,\Phi(a_{p-1}))
\]
for every $m\in M$ and $a_1,\ldots,a_{p-1}\in E_m$. In our case, the map $\Phi$ is  $\map{\prol{\pr_1}}{\prol[E]{(E^*\times_MB)}}{\prol[E]{E^*}}$, the prolongation of the map $\map{\pr_1}{E^*\times_MB}{E^*}$ (this last map f\/ibered over the identity in $M$), and $\sigma_H$ is a section along $\pr_1$. Therefore, $i_{\sigma_H}\Omega-dH$ is a section of the dual bundle to $\prol[E]{(E^*\times_MB)}$.

It is easy to see that the symplectic equation~\eqref{pontryagin} has a unique solution def\/ined on the following subset
\[
S_H=\set{(\mu,u)\in E^*\times_M B}{\pai{dH(\mu,u)}{V}=0\text{ for all }V\in\ker\prol{\pr_1}}.
\]
Therefore, it is necessary to perform a stabilization constraint algorithm to f\/ind the integral curves of $\sigma_H$ which are tangent to the constraint submanifold.

In local coordinates, the solution to the above symplectic equation is
\[
\sigma_H=\pd{H}{\mu_\alpha}\X_\alpha-\left[\rho^i_\alpha\pd{H}{x^i}+
\mu_\gamma C^\gamma_{\alpha\beta}\pd{H}{\mu_\beta}\right]\P^\alpha,
\]
def\/ined on the subset where
\[
\pd{H}{u^A}=0,
\]
and therefore the critical trajectories are the solution of the dif\/ferential-algebraic equations
\begin{gather*}
\dot{x}^i=\rho^i_\alpha\pd{H}{\mu_\alpha},\\
\dot{\mu}_\alpha
=-\left[\rho^i_\alpha\pd{H}{x^i}+\mu_\gamma C^\gamma_{\alpha\beta}\pd{H}{\mu_\beta} \right],\\
0=\pd{H}{u^A}.
\end{gather*}
Notice that $\pd{H}{\mu_\alpha}=\sigma^\alpha$.

One can easily see that whenever it is possible to write $\mu_\alpha=p_i\rho^i_\alpha$ then the above dif\/ferential equations reduce to the critical equations for the control system $Y=\rho(\sigma)$ on $TM$ and the index~$L$. Nevertheless it is not warranted that $\mu$ is of that form. For instance in the case of a~Lie algebra, the anchor vanishes, $\rho=0$, so that the factorization $\mu_\alpha=p_i\rho^i_\alpha$ will not be possible in general.

\subsubsection*{Reduction}
Consider two optimal control systems, with data $(B,E,M,\sigma,L)$ and $(B',E',M',\sigma',L')$. Let $\map{\Phi}{E}{E'}$ be a f\/iberwise bijective morphism of Lie algebroids over a map $\varphi$. Then $\map{\Phi_m}{E_m}{E'_{\varphi(m)}}$ is invertible for every $m\in M$, and we can consider the contragredient map $\map{\Phi^c}{E^*}{E^{'*}}$. This is the vector bundle map over $\varphi$ whose restriction to the f\/iber over $m\in M$ is $\map{\Phi_m^c=\Phi_m^{*-1}}{E^*_m}{E^{'*}_{\varphi(m)}}$, given by $\pai{\Phi^c(\mu)}{a}=\pai{\mu}{\Phi_m^{-1}(a)}$ for every $\mu\in E^*_m$ and every $a\in E'_{\varphi(m)}$.

If we further have a f\/ibered map $\map{\psi}{B}{B'}$ over the same map $\varphi$, then we can def\/ine the map $\map{\Psi}{E^*\times_MB}{E'{}^*\times_{M'}B'}$ by
\[
\Psi(\mu,u)=(\Phi^c(\mu),\psi(u)),\qquad (\mu,u)\in E^*\times_MB,
\]
and we have the following transformation properties.

\begin{proposition}
We have the following properties
\begin{enumerate}\itemsep=0pt
\item $(\prol[\Phi]{\Phi^c})\pb\Theta'=\Theta$.
\item $(\prol[\Phi]{\Phi^c})\pb\,\Omega'=\Omega$.
\item If $L'\circ\psi=L$, then $H'\circ\Psi=H$.
\end{enumerate}
\end{proposition}
\begin{proof}
For every $(\mu,b,v)\in\TEE^*$ we have
\begin{gather*}
\pai{(\prol[\Phi]{\Phi^c})\pb\Theta'}{(\mu,b,v)}
=\pai{\Theta'}{\prol[\Phi]{\Phi^c}(\mu,b,v)}
=\pai{\Theta'}{(\Phi^c(\mu),\Phi(b),T\Phi^c(v))}\\
\phantom{\pai{(\prol[\Phi]{\Phi^c})\pb\Theta'}{(\mu,b,v)}}{}=\pai{\Phi^c(\mu)}{\Phi(b)}
=\pai{\mu}{b}=\pai{\Theta}{(\mu,b,v)},
\end{gather*}
which proves the f\/irst.

The proof of the third is similar and will be omitted. For the second just take into account that $\prol[\Phi]{\Phi^c}$ is a morphism.
\end{proof}

By mean of standard manipulations one can easily show that
\[
(\prol[\Phi]{\Psi})\pb\bigl(i_{\sigma_{H'}}\Omega'-dH'\bigr)-\bigl(i_{\sigma_H}\Omega-dH\bigr)=
\Omega'\Bigl(\sigma_{H'}\circ\Psi-\prol[\Phi]{\Phi^c}\circ\sigma_H, \prol[\Phi]{\Psi}(\,\,\cdot\,\,)\Bigr),
\]
from where the following theorem, which establishes the relation between critical trajectories of the two related optimal control problems, readily follows.
\begin{theorem}
Let $\map{\psi}{B}{B'}$ and $\map{\Phi}{E}{E'}$ be fibered maps over the same map $\map{\varphi}{M}{M'}$, and assume that $\psi$ is fiberwise submersive and $\Phi$ is a morphism of Lie algebroids which is fiberwise bijective. Let $L$ be an index function on $B'$ and $L'$ be an index function on $B'$ such that $L=L'\circ\psi$ and let $\sigma_H$ and $\sigma_{H'}$ the corresponding critical sections. Then we have that $\Psi(S_H)\subset S_{H'}$ and
\[
\prol[\Phi]{\Phi^c}\circ\sigma_H=\sigma_{H'}\circ\Psi
\]
on the subset $S_H$.

As a consequence, the image under $\Psi$ of any critical trajectory for the index $L$ is a critical trajectory for the index $L'$.
\end{theorem}

As an application of the above result we can consider the case of reduction by a symmetry group (with a free and proper action). Indeed, applying this result to $E=TQ$, $B=B$, $M=Q$ and $E'=TQ/G$, $B'=B/G$, $M'=Q/G$, with $\psi(b)=[b]$, $\Phi(v)=[v]$, $\varphi(q)=[q]$, the quotient maps and index $L$, $L'([b])=L(b)$ (so that $L=L'\circ\psi$) we have that any critical trajectory for $L$ in $Q$ is a critical trajectory for $L'$ in the reduced space $M'$.

Finally notice that the case of Hamiltonian mechanics corresponds to $B=E$ and $\sigma=\id$, and hence the set $S_H$ is the graph of the Legendre transform. Therefore, the above results about optimal control produce also results about reduction of Hamiltonian systems. It would be nice to have a similar result for f\/iberwise surjective (no necessarily f\/iberwise invertible) morphisms.

\subsection*{Acknowledgements}
Partial f\/inancial support from MEC-DGI (Spain) grants BFM~2003-02532 and MTM2006-10531 is acknowledged.

\pdfbookmark[1]{References}{ref}
\LastPageEnding

\end{document}